\documentclass[aps,pra,reprint,superscriptaddress]{revtex4-2}

\usepackage{graphicx} 
\usepackage{amsmath,amssymb,bm} 

\begin{document}

\title{Observation of Localization Using a Noisy Quantum Computer}

\author{Kazue Kudo}
\email[]{kudo@is.ocha.ac.jp}
\affiliation{Department of Computer Science, Ochanomizu University, Tokyo 112-8610, Japan}
\affiliation{Graduate School of Information Sciences, Tohoku University, Sendai 980-8579, Japan}

\date{\today}

\begin{abstract}
Quantum dynamics in a strongly disordered quantum many-body system show localization properties.
The initial state memory is maintained owing to slow relaxation when the system is in the localized regime. 
This work demonstrates how localization can be observed using a noisy quantum computer by evaluating the magnetization and twist overlap in a quantum spin chain after short-time evolution.
 The quantities obtained from quantum-circuit simulation and real-device computation show their apparent dependence on disorder strength, although real-device computation suffers from noise-induced errors significantly.
 Using the exact diagonalization of the Hamiltonian, we analyze how noise-induced errors affect those quantities.
 The analysis also suggests how the twist overlap can reflect the information on the eigenstates of the Hamiltonian.
\end{abstract}


\maketitle

\section{Introduction}

Localization in quantum systems has been extensively investigated across a wide variety of contexts.
One particular area of interest is many-body localization (MBL), which is eigenstate localization in strongly disordered interacting systems.
MBL can be regarded as an extended concept of Anderson localization.
Despite decades of investigation, MBL is still attracting considerable attention~\cite{parameswaran2018,abanin2019,gopalakrishnan2020,tikhonov2021}.
Many numerical studies of the MBL transition using various quantities derived from eigenstates~\cite{pal2010,serbyn2013b,serbyn2016,khemani2016,khemani2017a,khemani2017b,bera2015,bera2016,kjall2014,luitz2015,hopjan2020,gray2018,kudo2018} and quantum dynamics~\cite{bardarson2012,serbyn2013a,enss2017} have demonstrated features of a continuous transition from the thermal to localized phases.
However, the exponents of finite-size scaling analysis in quenched disordered systems violate the Harris criterion~\cite{harris1974,chayes1986}, indicating that the system sizes used in those numerical studies are too small to properly analyze the MBL transition.  
In other words, the transition observed in the numerical works of finite-size systems is not a true phase transition but rather a crossover. 
Although the existence of the true MBL phase is controversial, the MBL transition point is located at a much stronger disorder if it exists~\cite{imbrie2016,sierant2022,morningstar2022,sels2022}.
In this paper, we consider a finite-size system to be in the localized regime where it displays certain MBL properties.

Numerous experimental studies have shown the crossover from the thermal to localized regimes~\cite{schreiber2015,smith2016,bordia2016,bordia2017,luschen2017a,luschen2017b,xu2018,kohlert2019,rubio2019,gong2021,filho2022,chiaro2022,liu2023}.
Methods based on the memory effect are commonly used to detect MBL in experiments, as slow relaxation in the localized regime results in the initial state remaining almost unchanged for a long time.
Experimental investigations often focus on simple quantities such as imbalance and magnetization, which can be easily obtained by qubit measurements. 
However, these quantities do not provide much information on eigenstates of the system Hamiltonian.
Although tomographic techniques can provide information on eigenstates, they require significant effort.
In contrast, twist overlap, which can also be easily obtained from qubit measurements, exhibits distinct behavior beyond the memory effect and can reflect the information on eigenstates~\cite{kudo2022}:
The twist overlap obtained from quantum dynamics behaves similarly to that evaluated using eigenstates.

The objective of this study is to investigate the observation of MBL using a noisy quantum computer.
The effect of noise varies depending on the type of observable.
In this work, we aim to provide theoretical and numerical explanations of why noise-induced errors have different effects on different observables.
In particular, we focus on localization detection based on quantum dynamics and demonstrate the disorder dependence of both the magnetization and twist overlap.
A comparison of results obtained from quantum-circuit simulation and real-device computation provides insights into noise-induced errors.
 Our findings reveal that the twist overlap is more vulnerable to noise-induced errors than the magnetization.
Furthermore, we analyze the results, especially in the localized regime, using the exact diagonalization of the Hamiltonian.
The analysis suggests why the twist overlap is more susceptible to noise and how it can reflect the information on eigenstates.

In this study, we employ the method of localization detection using a gate-model quantum computer, inspired by experiments of localization detection using a quantum annealer~\cite{filho2022}.
When using the quantum annealer, it is difficult to adjust interaction strength and local field separately.
In contrast, a gate-model quantum device allows the independent control of the parameters for interaction strength and local field, enabling the separate control of disorder strength.
However, the quantum-circuit computation of time evolution suffers from approximation error owing to Suzuki--Trotter decomposition~\cite{hatano2005,childs2021}.
Increasing the Trotter number reduces the approximation error, but it also increases the number of quantum gates, thereby increasing gate errors during real-device computation.
Thus, the balance between the approximation and gate errors is critical in a noisy quantum device.
To reduce noise-induced errors, we focus on short-time behavior instead of long-time evolution.
Although approximate quantum circuits can further reduce noise-induced errors~\cite{jaderberg2020,jaderberg2022,liu2023}, we employ a straightforward method based on the Suzuki--Trotter decomposition in this work because results obtained using this simple method are preferable for error analysis.

\section{\label{model} Model and Methods}

\subsection{Hamiltonian and Observables}

The Hamiltonian of a one-dimensional transverse Ising model with local random fields consists of the Ising Hamiltonian $H_{\rm I}$ and the transverse-field Hamiltonian $H_{\rm TF}$: $H= H_{\rm I} + H_{\rm TF}$, where
\begin{align}
  H_{\rm I} &= \sum_{j=1}^{L-1}J\sigma^z_j\sigma^z_{j+1}
  + \sum_{j=1}^L h_j\sigma^z_j,\\
 H_{\rm TF} &= - \sum_{j=1}^L \Gamma\sigma^x_j.
\end{align}
Here, $L$ is the system size, and $\sigma_j^x$ and $\sigma_j^z$ represent the Pauli operators of components $x$ and $z$, respectively, at site $j$.
The interaction is antiferromagnetic; hence, the parameter $J$ is positive.
The strength of the transverse field is denoted by $\Gamma$ and is also taken as positive.
Throughout this study, these parameters are fixed to $J=\Gamma=1$.
The local fields $h_j$ are uniformly distributed random numbers in the interval $[-w, w]$, where $w$ denotes the disorder strength. 

We observe the magnetization and twist overlap at the end of the time evolution, namely, at the final time $t=T_{\rm fin}$.
Both observables are easily obtained by measuring each qubit in a quantum device.
These observables averaged over different disorder realizations serve as probes for localization~\cite{kudo2022}.
In this study, the initial state is given as the all-spin-up state.

The magnetization in the $z$ direction is defined as $M_z=\langle\psi|\hat{S}^z|\psi\rangle$, where $|\psi\rangle$ is a wavefunction, and
\begin{align}
 \hat{S}^z=\sum_{j=1}^L\sigma_j^z.
\end{align}
Since $\hat{S}^z$ is diagonal in the computational basis, its eigenstates are given by the computational basis states $|k\rangle$, which satisfy
\begin{align}
 \hat{S}^z|k\rangle = m_k|k\rangle,
 \quad
 m_k=\sum_{j=1}^L s_j^{(k)}.
\end{align}
Here, $m_k$ is the eigenvalue corresponding to $|k\rangle$.
The spin configuration corresponding to $|k\rangle$ is described by $\bm{s}^{(k)}\in\{-1,1\}^L$, whose $j$th element is $s_j^{(k)}$.
In the quantum-circuit simulation and real-device computation, the magnetization is expressed as $M_z=\sum_{k}p_km_k$, where $p_k$ is the experimental probability associated with the spin configuration $\bm{s}^{(k)}$.

The magnetization, which is initially equal to the system size $L$, relaxes to zero because of antiferromagnetic interactions in a weak-disorder case.
In contrast, the magnetization in a strong-disorder case is expected to retain its initial value owing to the memory effect.

The twist overlap $z=\langle\psi|\hat{U}_{\rm twist}|\psi\rangle$ is the overlap between a state $|\psi\rangle$ and its twisted state $\hat{U}_{\rm twist}|\psi\rangle$.
Here, $\hat{U}_{\rm twist}$ is the twist operator defined by
\begin{equation}
 \hat{U}_{\rm twist}=\exp\left[ i\sum_{j=1}^L\theta_j\sigma^z_j/2 \right],
\end{equation}
where $\theta_j=2\pi j/L$.
Since the twist operator is also diagonal in the computational basis, we have
\begin{align}
 \hat{U}_{\rm twist}|k\rangle = \exp(iu_k)|k\rangle,
 \quad
 u_k=\frac{\pi}{L}\sum_{j=1}^L js_j^{(k)}.
\end{align}
In the quantum-circuit simulation and real-device computation, the twist overlap is given by $z=\sum_{k}p_k\exp(iu_k)$.
Consequently, both the twist overlap and the magnetization are derived from an identical set of measurement outcomes.

The twist operator generates a spin-wave-like excitation through the rotation of spins around the $z$-axis at angles $\theta_j$~\cite{nakamura2002,kutsuzawa2022}.
When $|\psi\rangle$ is a thermal eigenstate, the twist overlap vanishes because the twisted state with a spin-wave-like excitation is orthogonal to the original state.
However, the long-wavelength perturbation given by the twist operator has little effect on localized eigenstates, leading to a finite twist overlap.
The twist overlap can also be evaluated using a wavefunction that is a linear combination of eigenstates.

\subsection{Time Evolution}

Time evolution is based on the Schr{\"o}dinger equation
\begin{equation}
 i\frac{d}{dt}|\psi(t)\rangle
=H|\psi(t)\rangle,
\end{equation}
whose solution is represented as
\begin{equation}
 |\psi(t)\rangle = e^{-iHt}|\psi_0\rangle,
\label{eq:psi1}
\end{equation}
where $|\psi_0\rangle$ denotes the initial state.
The wavefunction $|\psi(t)\rangle$ at any time can be obtained through the exact diagonalization of the Hamiltonian.
Writing the initial state as a liner combination $|\psi_0\rangle=\sum_k c_k|\phi_k\rangle$ of the eigenstates $|\phi_k\rangle$ yields
\begin{equation}
 |\psi(t)\rangle=\sum_{k=1}^{2^L}c_ke^{-iE_kt}|\phi_k\rangle,
\label{eq:psi.2}
\end{equation}
where $E_k$ is the eigenenergy corresponding to $|\phi_k\rangle$.
We here refer to the time dependence of wavefunctions obtained from the exact diagonalization as the exact time evolution. 

For quantum simulations using quantum circuits, the time evolution operator should be approximated using product formulas~\cite{hatano2005,childs2021}.
Using the first-order Suzuki--Trotter decomposition, we have
\begin{align}
e^{-iH\tau} = e^{-i(H_{\rm I}+H_{\rm TI})\tau}
=\left( e^{-iH_{\rm I}\delta}e^{-iH_{\rm TI}\delta}\right)^m,
\end{align}
where $\delta=\tau/m$ and $m$ is called the Trotter number or Trotter steps.
Similarly, the second-order Suzuki--Trotter decomposition leads to
\begin{align}
e^{-iH\tau} = \left( e^{-iH_{\rm TI}\delta/2}e^{-iH_{\rm I}\delta}
e^{-iH_{\rm TI}\delta/2}\right)^m.
\end{align}
The Trotter error for the $p$th-order decomposition scales as $O(\delta^{p+1})$.
Single-qubit rotation gates applied to the $j$th qubit denoted by $R^X_j(\theta)$ and $R^Z_j(\theta)$ implement the unitary operators $e^{-i(\theta/2)\sigma_j^x}$ and $e^{-i(\theta/2)\sigma_j^z}$, respectively.
Similarly, two-qubit $ZZ$ rotation gates implement operators for interaction terms.

\subsection{Real-Device Computation}

We utilized \texttt{ibmq\_manila}, one of the IBM Quantum Falcon Processors, to perform computation on a real quantum computer.
The machine consists of five qubits arranged in a linear structure.

The straightforward execution of quantum circuits on noisy quantum hardware causes errors.
To reduce noise-induced errors without relying on fault tolerance strategies, we utilized a measurement mitigation technique called matrix-free measurement mitigation (M3), which is implemented through the \texttt{mthree} Qiskit package~\cite{mthree}. 
M3 computes corrected measurement probabilities within a reduced subspace, using either the lower-upper (LU) decomposition or a preconditioned iterative method.

We evaluated observables by measurements with 1024 shots per circuit in both quantum-circuit simulation and real-device computation.

\section{Disorder Dependence}
\label{sec:depend}

We measured the magnetization and twist overlap at the end of time evolution,
which we fixed at $T_{\rm fin}=1.5$ in this study.
This time is sufficiently long, especially for cases with middle to strong disorders, as supported by the time evolution results presented in Sect.~\ref{sec:time}.
We set the system size to $L=5$ for quantum-circuit simulation, real-device computation, and exact time evolution to facilitate comparison.

\begin{figure}[tb]
 \includegraphics[width=4.2cm]{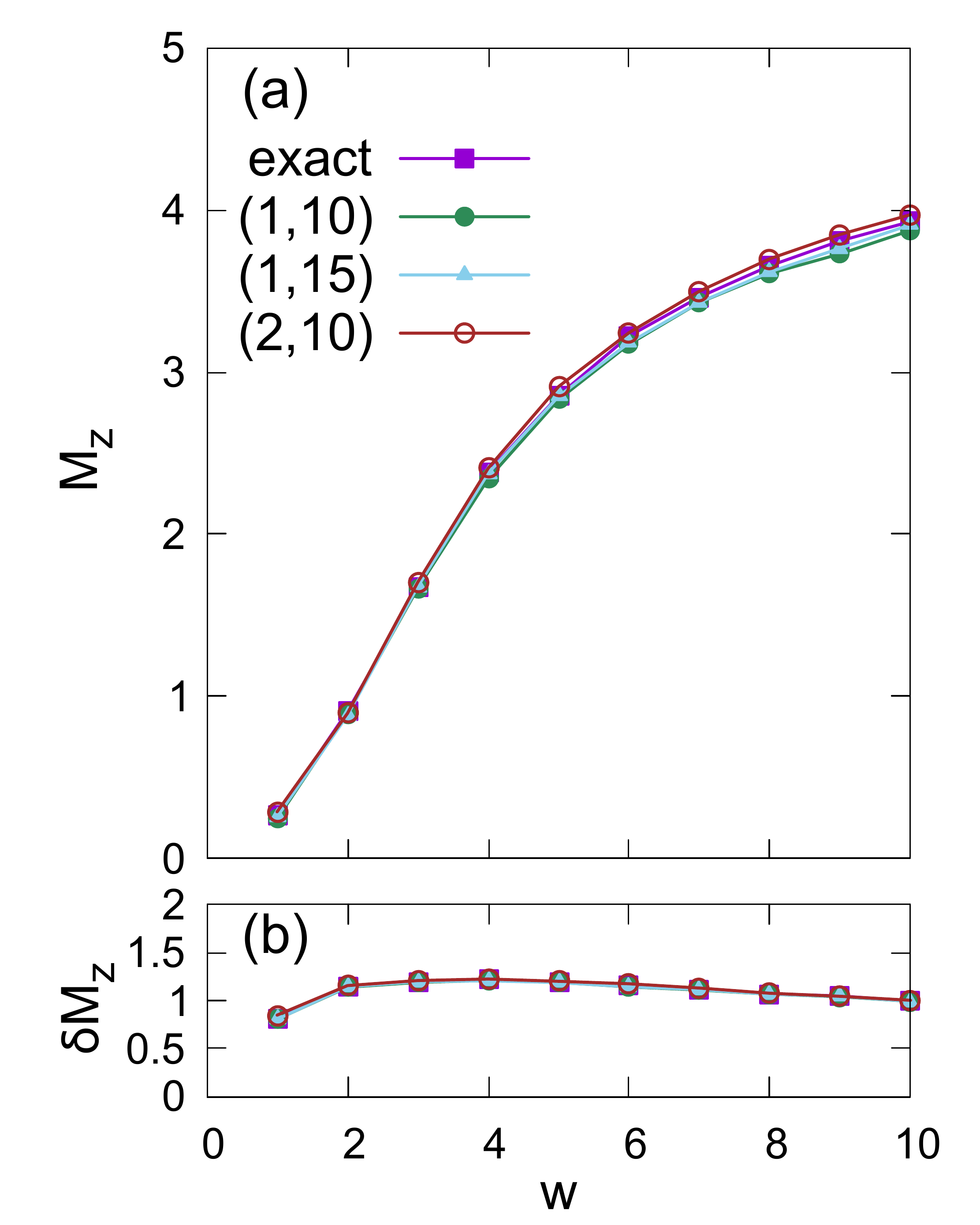}
 \includegraphics[width=4.2cm]{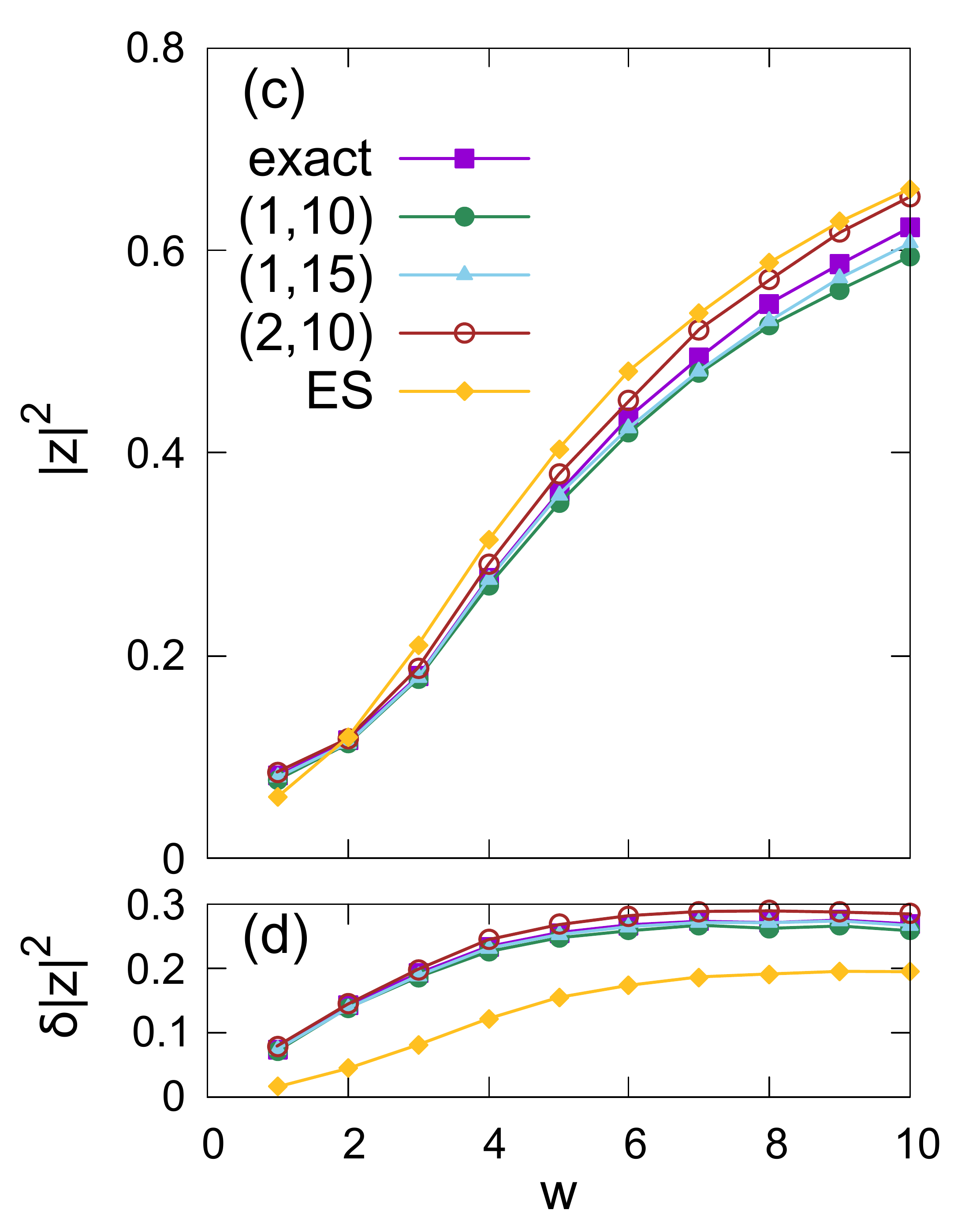}%
 \caption{\label{fig:w_mzz2s} (Color online) Disorder dependence of (a,b) the magnetization in the $z$-direction $M_z$ and (c,d) the absolute square of the twist overlap $|z|^2$.
 Panels (a) and (c) show the average, and panels (b) and (d) show the standard deviation.
 The data obtained from the exact time evolution are labeled ``exact''.
 The simulation results obtained from the $p$th-order Suzuki--Trotter decomposition with the Trotter number $m$ are labeled $(p,m)$.
 The data evaluated with eigenstates of the Hamiltonian are labeled ``ES''. 
}
\end{figure}

The results of quantum-circuit simulations, computed using a classical computer, are shown in Fig.~\ref{fig:w_mzz2s}.
The simulation results labeled $(p,m)$ are obtained using the $p$th-order Suzuki--Trotter decomposition with the Trotter number $m$.
The data are averaged over $10^4$ disorder realization samples.
The results labeled ``exact'' are obtained from the exact time evolution.
The disorder dependence of $M_z$ in the simulations is almost identical to that in the exact computation.

The twist overlap simulation results are compared with those obtained from the exact time evolution and the ones evaluated with eigenstates (labeled ``ES'') of the Hamiltonian.
The twist overlap evaluated with eigenstates in each realization is calculated and averaged over 16 eigenstates around the center of the energy spectrum.
The average of $|z|^2$ in the simulations shows a trend similar to the results obtained from the exact time evolution and the ones evaluated with eigenstates. 
However, the standard deviation $\delta|z|^2$ for the results obtained from  eigenstates is smaller than the others.
The simulations based on the first-order Suzuki--Trotter decomposition yields a smaller $|z|^2$ than the exact time evolution, whereas the second-order decomposition leads to a larger $|z|^2$.

\begin{figure}[tb]
 \includegraphics[width=4.2cm]{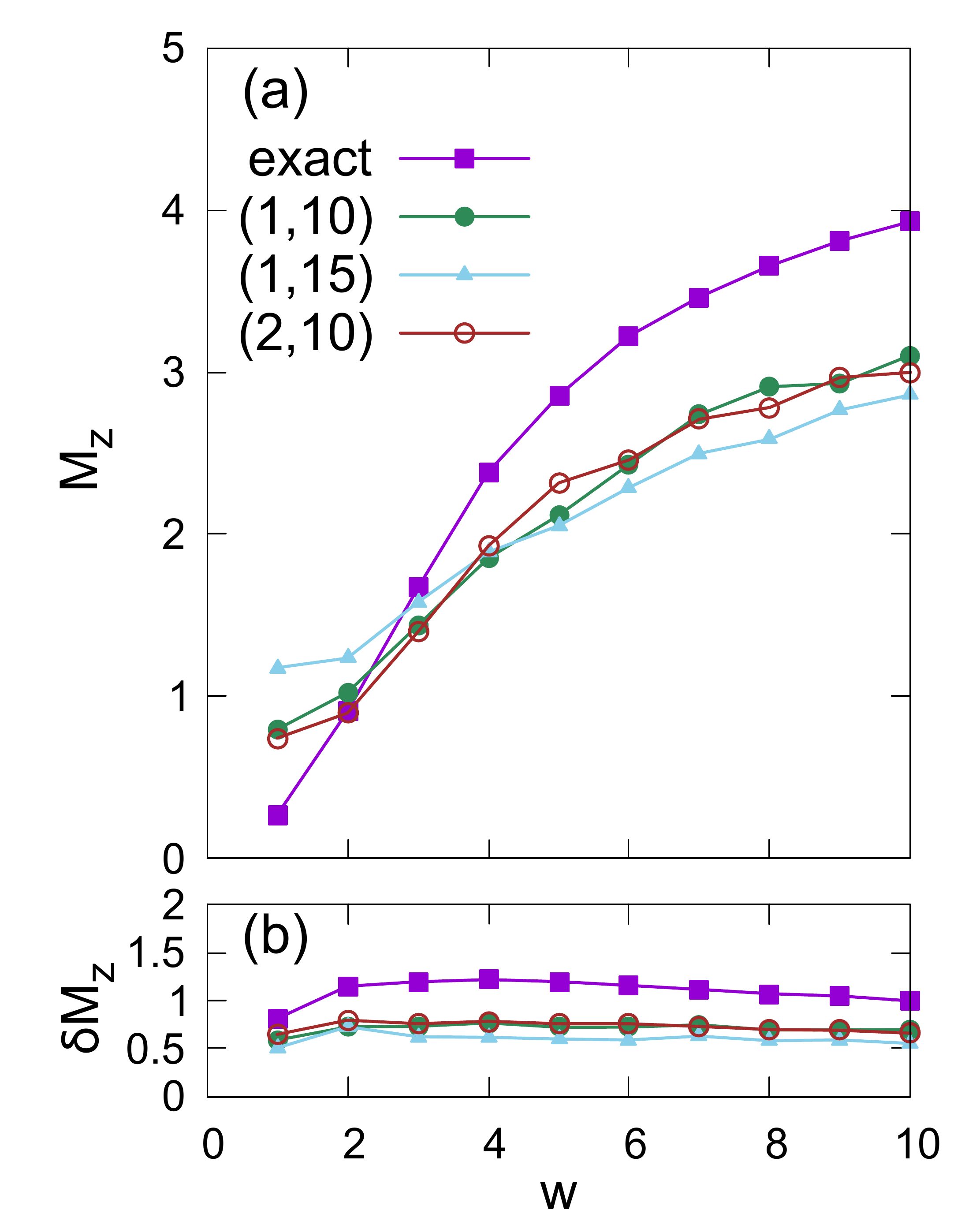}
 \includegraphics[width=4.2cm]{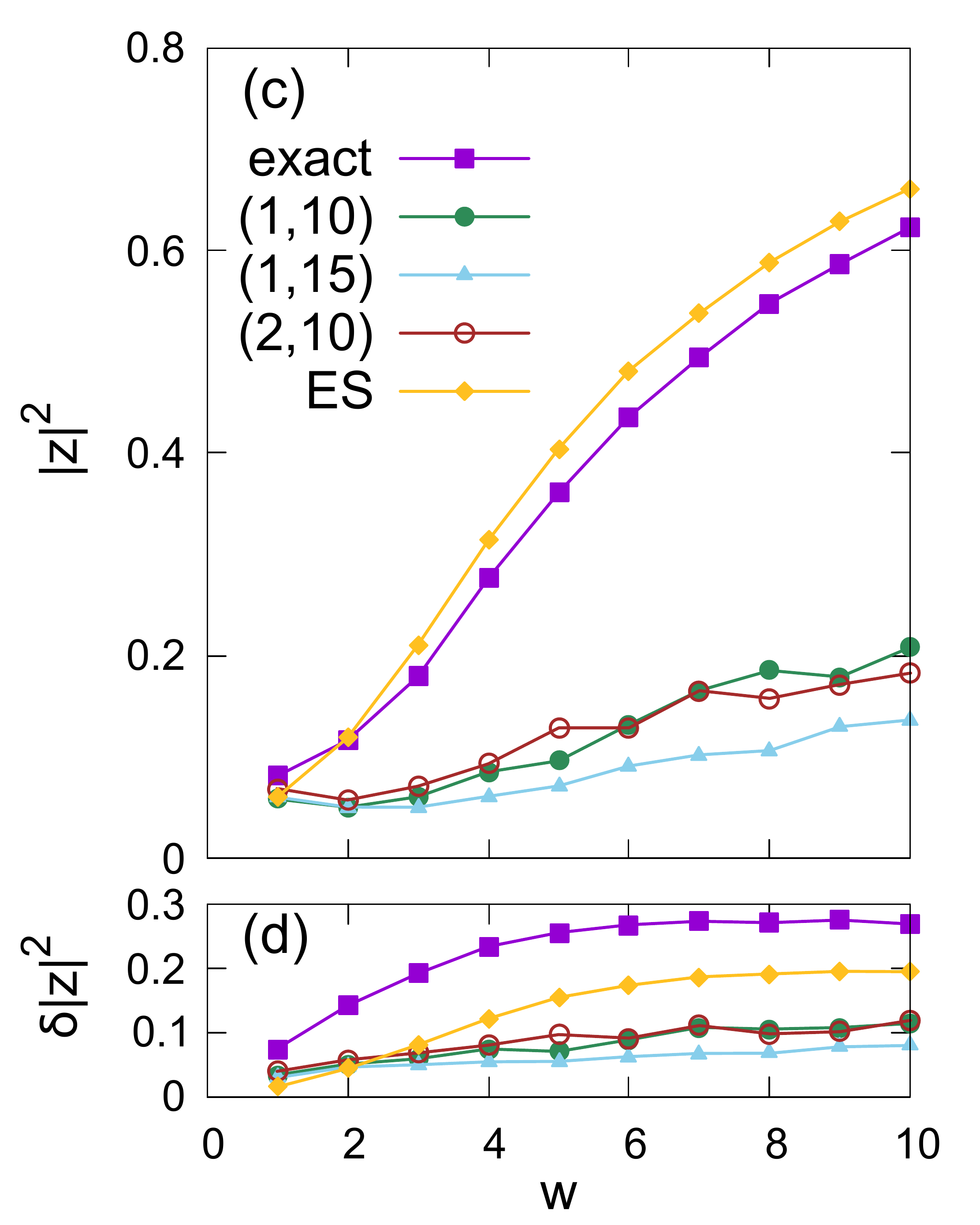}%
\caption{\label{fig:w_mzz2r} (Color online) Disorder dependence of (a,b) the magnetization in the $z$-direction $M_z$ and (c,d) the absolute square of the twist overlap $|z|^2$.
 Panels (a) and (c) show the average, and panels (b) and (d) show the standard deviation.
 The real-device computation results for the $p$th-order Suzuki--Trotter decomposition with the Trotter number $m$ are labeled $(p,m)$.
The data labeled ``exact'' and ``ES'' are the same as those in Fig.~\ref{fig:w_mzz2s}. 
}
\end{figure}

The real-device computation results shown in Fig.~\ref{fig:w_mzz2r} indicate that noise-induced errors are significant.
The data labeled $(p,m)$ are computed using the $p$th-order Suzuki--Trotter decomposition with the Trotter number $m$ and averaged over $10^3$ disorder realization samples.
The magnetization obtained from real-device computation increases with disorder strength. 
However, the average $M_z$ is smaller than the result obtained from the exact time evolution, particularly in the large-$w$ region.
The average and standard deviation of $|z|^2$ obtained from real-device computation also increase with disorder strength, although they are much smaller than those obtained from the exact time evolution.
The errors in $M_z$ and $|z|^2$ are more significant for $m=15$ than for $m=10$, whereas the difference between $p=1$ and $p=2$ is negligible.

\section{Time Evolution}
\label{sec:time}

\begin{figure}[tb]
\centering
  \includegraphics[width=8.5cm]{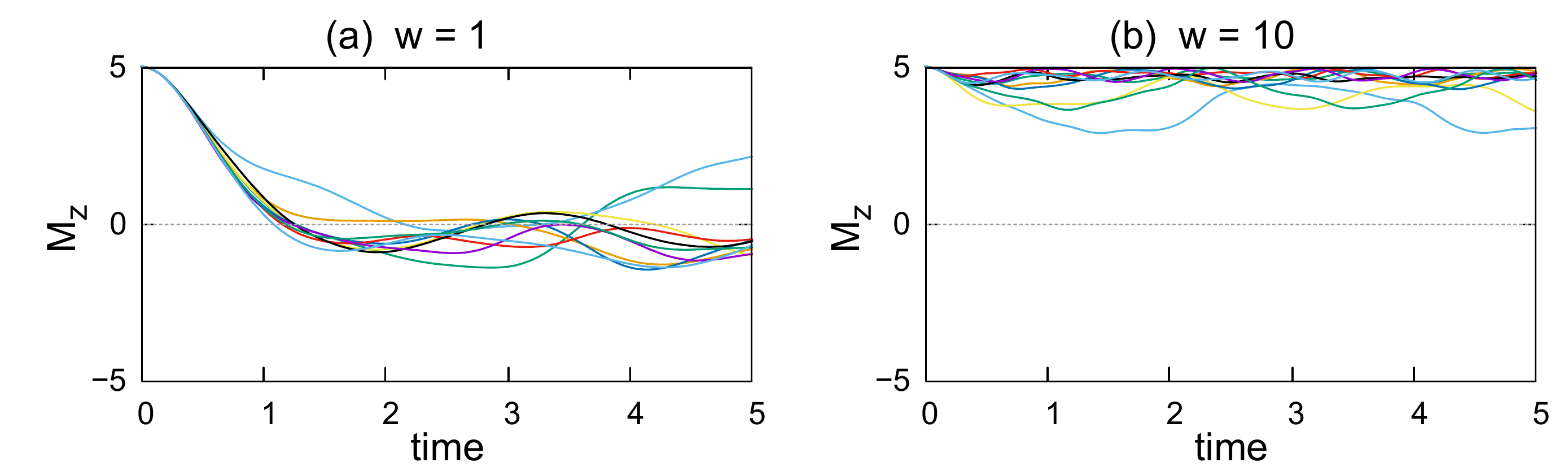}\\
  \includegraphics[width=8.5cm]{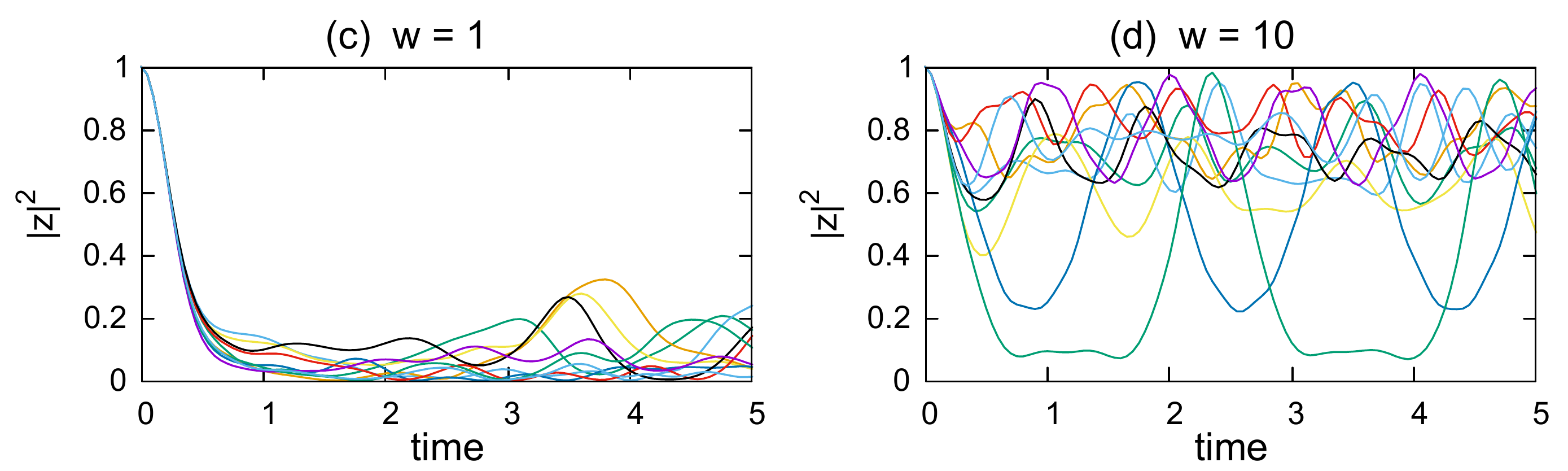}%
\caption{\label{fig:time_evolv} (Color online) Time evolution of (a,b) the magnetization in the $z$-direction $M_z$ and (c,d) the absolute square of the twist overlap $|z|^2$ for a system size of $L=5$. 
Each panel displays ten different samples.
 Panels (a) and (c) share the same samples of wavefunctions, and the same applies to panels (b) and (d).}
\end{figure}

Figure~\ref{fig:time_evolv} shows the time evolution of the magnetization in the $z$-direction ($M_z$) and the absolute square of the twist overlap ($|z|^2$) for ten different disorder realization samples, calculated using the exact time evolution for a system with $L=5$.
Figures~\ref{fig:time_evolv}(a) and \ref{fig:time_evolv}(c) share the same samples of wavefunctions, implying that each sample of $M_z$ is obtained from the same realization as the corresponding sample of $|z|^2$.
The same applies to Figs.~\ref{fig:time_evolv}(b) and \ref{fig:time_evolv}(d).
In a weak disorder case ($w=1$), the initial decay of $M_z$ ceases around $t\simeq 1$, whereas that of $|z|^2$ does before $t=1$.  
These results suggest that the final time should be $T_{\rm fin}>1$.
To reduce noise-induced errors, a short $T_{\rm fin}$ is preferable.
Thus, we set $T_{\rm fin}=1.5$ for computation in Sect.~\ref{sec:depend}.

Although both the magnetization and the twist overlap tend to relax to zero in the thermal regime ($w=1$), they exhibit different behaviors in the localized regime ($w=10$).
The magnetization in the localized regime fluctuates around the initial value owing to the memory effect, whereas $|z|^2$ does not maintain the initial value.
The time average of $|z|^2$ is less than unity, although it is still much larger than that in the thermal regime, and the variation in time average is notable. 
Moreover, some samples of $|z|^2$ show large-amplitude oscillations in the localized regime.
These properties are reflected in the large variance of the twist overlap shown in Fig.~\ref{fig:w_mzz2s}(c).

To understand the difference, we theoretically analyze the time evolution of the system, focusing on the localized regime.
Using Eq.~\eqref{eq:psi.2}, we describe the time dependence of the magnetization and twist overlap as follows:
\begin{align}
 M_z(t) &= \sum_k|c_k|^2 \langle\phi_k|\hat{S}^z|\phi_k\rangle
 \nonumber\\
&\quad +\sum_{k\neq l}c_kc_l^*\langle\phi_l|\hat{S}^z|\phi_k\rangle e^{-i(E_k-E_l)t},
\label{eq:Mz.t}\\
z(t) &= \sum_k|c_k|^2 \langle\phi_k|\hat{U}_{\rm twist}|\phi_k\rangle
 \nonumber\\
&\quad+\sum_{k\neq l}c_kc_l^*\langle\phi_l|\hat{U}_{\rm twist}|\phi_k\rangle
e^{-i(E_k-E_l)t}.
\label{eq:z.t}
\end{align}
The first terms on the right-hand side of Eqs.~\eqref{eq:Mz.t} and \eqref{eq:z.t} represent long-time averages, whereas the second terms correspond to time-dependent parts.

In the MBL phase, the eigenstates of the Hamiltonian are approximately expressed as product states of local integrals of motion (LIOMs)~\cite{huse2014,serbyn2013b,chiaro2022}.
Assuming that these states are also sufficiently close to the computational basis in the localized regime considered here, we describe the eigenstate $|\phi_k\rangle$ corresponding to the basis state $|k\rangle$ as a linear combination of $|k\rangle$ and a state perpendicular to $|k\rangle$, i.e.,
\begin{align}
  |\phi_k\rangle = b_k|k\rangle + |\alpha_k\rangle,
\label{eq:decomp}
\end{align}
where $\langle k|\alpha_k\rangle=0$.
We also assume that $\left|\langle\alpha_k|\alpha_l\rangle\right|\ll 1$ for $l\neq k$.

Using Eqs.~\eqref{eq:Mz.t} and \eqref{eq:decomp} and assuming $\left|\langle\alpha_k|\alpha_l\rangle\right|\ll 1$, we approximate the time dependence of magnetization as
\begin{align}
 M_z(t) &\simeq \sum_k|c_k|^2|b_k|^2 m_k +2\sum_km_k g_k(t),
\label{eq:Mz.t1}
\end{align}
where
\begin{align}
 g_k(t) &= \sum_{l (\neq k)}|A_{kl}|\cos[(E_k-E_l)t-\beta_{kl}],
\label{eq:g_k}
\end{align}
with $A_{kl}=c_kc_l^*b_k$.
Here, $\beta_{kl}$ is defined through 
$\tan\beta_{kl} = (\mathrm{Im}\; A_{kl})/(\mathrm{Re}\; A_{kl})$.
Similarly, using Eqs.~\eqref{eq:z.t} and \eqref{eq:decomp} and $\left|\langle\alpha_k|\alpha_l\rangle\right|\ll 1$, we approximate $|z(t)|^2$ as 
 \begin{align}
  |z(t)|^2 &\simeq |z_0|^2 + 4|z_0|\sum_k\cos(u_k-\beta_0)g_k(t)
  \nonumber\\
  &\quad +\left|\sum_k e^{iu_k}g_k(t)\right|^2,
\label{eq:z2.t}
 \end{align}
where $z_0= \sum_k|c_k|^2 \langle\phi_k|\hat{U}_{\rm twist}|\phi_k\rangle$
and $\beta_0$ is defined through
$\tan\beta_0 = (\mathrm{Im}\; z_0)/(\mathrm{Re}\; z_0)$.

The long-time averages of $M_z$ and $|z|^2$ in the localized regime are approximated as follows:
\begin{align}
  \overline{M_z(t)}&\simeq \sum_k|c_k|^2|b_k|^2 m_k,
\label{eq:av_mz}\\
 \overline{|z(t)|^2} 
&\simeq \sum_k|c_k|^4|b_k|^4.
\label{eq:av_z2}
\end{align}
Here, we have neglected the terms related to $|A_{kl}|^2$, which are assumed to be negligible for $k\neq l$.
Since $\overline{M_z(t)}\simeq L$ in the localized regime, as shown in Fig.~\ref{fig:time_evolv}(b), we can assume that $|c_k|^2\approx 1$ and $|b_k|^2\approx 1$ at $k$, where $m_k= L$.
Additionally, since $|c_k|^4<|c_k|^2$ and $|b_k|^4<|b_k|^2$, we can also understand why $\overline{|z(t)|^2}$ is slightly smaller than unity.

\begin{figure}[tb]
\centering
  \includegraphics[width=8.5cm]{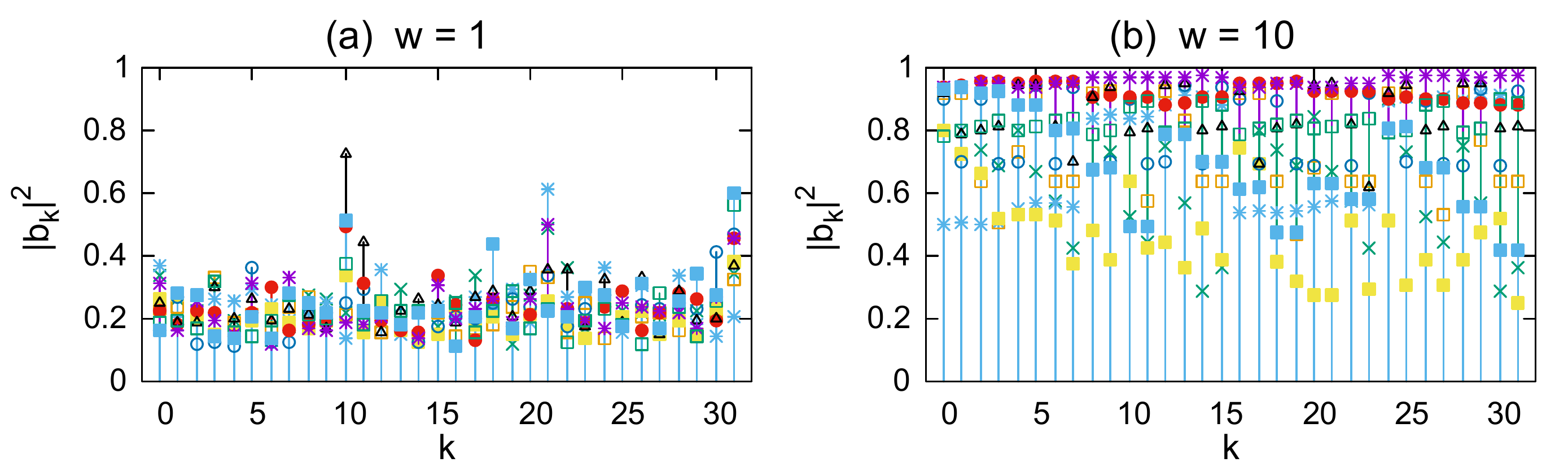}\\
  \includegraphics[width=8.5cm]{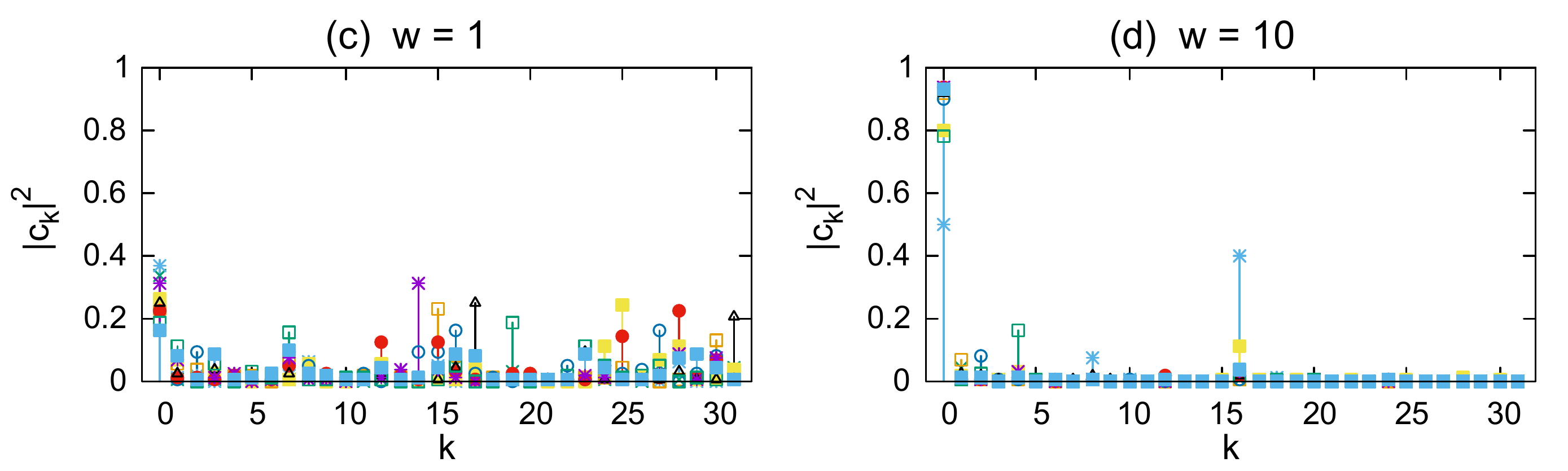}%
 \caption{\label{fig:bkck} (Color online) (a,b) Plots of the overlap $|b_k|^2$ between the basis state $|k\rangle$ and the corresponding eigenstate $|\phi_k\rangle$ of the Hamiltonian.
 (c,d) Plots of the overlap $|c_k|^2$ between the eigenstate $|\phi_k\rangle$ and the initial state $|\psi_0\rangle$.
Each panel displays ten samples that are obtained from the same samples of wavefunctions as in Fig.~\ref{fig:time_evolv}.
}
\end{figure}

Our assumptions in the localized regime are supported numerically.
Figure~\ref{fig:bkck} shows the overlap $|b_k|^2=|\langle k|\phi_k\rangle|^2$ between the basis state $|k\rangle$ and the corresponding eigenstate $|\phi_k\rangle$ of the Hamiltonian, 
as well as the overlap $|c_k|^2=|\langle\phi_k|\psi_0\rangle|^2$ between the eigenstate $|\phi_k\rangle$ and the initial state $|\psi_0\rangle$, for $w=1$ and $w=10$.
Here, we index the eigenstate $|\phi_k\rangle$ of the Hamiltonian as
$k=\mathrm{argmax}_i |\langle k|\phi_i\rangle|^2$.
Note that $|b_0|^2=|c_0|^2$ because the initial state is equivalent to the basis state of $k=0$.
In several samples for $w=10$, we observe that $|b_k|^2\simeq 1$, which is consistent with the assumption that the eigenstates are close to the computational basis in the localized regime.

The distributions of $|b_k|^2$ and $|c_k|^2$ also explain the behavior of time averages in the localized regime.
Whereas $|b_k|^2\simeq 1$ in typical samples for $w=10$, $|b_k|^2\sim 0.5$ in some samples, although $|b_k|^2$ is still mostly larger than that for $w=1$.
The variation in $|b_k|^2$ is related to that in $|z|^2$.
Additionally, $|c_k|^2$ has the largest peak at $k=0$, which corresponds to the initial state, and for most samples with $w=10$, $|c_0|^2\approx 1$ and $|c_k|^2\ll 1$ for $k\neq 0$.
These observations explain why $\overline{M_z(t)}\simeq m_0 =L$ and $\overline{|z(t)|^2}\simeq 1$ for typical samples in the localized regime.

Here, we analyze the time-dependent parts of the magnetization and twist overlap given by $f_{\rm mg}(t)$ and $f_{\rm tw}(t)$, respectively, which are expressed as
\begin{align}
  f_{\rm mg}(t) &= 2\sum_km_k  g_k(t),
\label{eq:f_mg}\\
 f_{\rm tw}(t) &= 4|z_0|\sum_k\cos(u_k-\beta_0) g_k(t).
\label{eq:f_tw}
\end{align}
We have neglected the terms including $|A_{kl}|^2$, which are assumed to be negligible for $k\neq l$.
Since both $f_{\rm mg}(t)$ and $f_{\rm tw}(t)$ depend on the same time-dependent function $g_k(t)$, they have the same period and phase of oscillation.
However, as seen in Figs.~\ref{fig:time_evolv}(b) and \ref{fig:time_evolv}(d), their oscillation amplitudes differ considerably in the localized regime.
The difference originates from the factors $m_k$ and $\cos(u_k-\beta_0)$.
As $m_k=0$ for many basis states, only a few elements of $g_k$ contribute to $f_{\rm mg}$.
The lack of contributions from the state with $m_k=0$ explains the reason for the relatively small amplitudes of magnetization oscillation seen in Fig.~\ref{fig:time_evolv}(b).
In contrast, $\cos(u_k-\beta_0)$ can take various values within the range $[-1,1]$.
Most elements of $g_k$ contribute to $f_{\rm tw}$, interfering constructively or destructively.
The variety in this factor explains the various amplitudes of $|z(t)|^2$ in Fig.~\ref{fig:time_evolv}(d).

The above theoretical and numerical explanations are based on the exact diagonalization of the Hamiltonian.
They capture the crucial aspects of the difference between the magnetization and the twist overlap.

\section{Discussion}

\subsection{Trotter Error and Noise-Induced Errors}

The numbers of single- and two-qubit gates used in real-device computation depend on the basis gates supported by the specific quantum device employed.
In the real device used in this work, an R$_{ZZ}$ gate is implemented as an R$_Z$ gate wrapped by a pair of CNOT gates, whereas an R$_X$ gate is substituted with a combination of four single-qubit gates.
In total, for $n$ qubits and $m$ Trotter steps, the numbers of single-qubit gates in the first- and second-order decompositions are $(6n-1)m$ and $(6n-1)m+5n$, respectively.
The number of two-qubit gates is $2(n-1)m$ for both the first- and second-order decompositions.

In the real device used in this work, single- and two-qubit gates are calibrated frequently.
The errors of CNOT and $\sqrt{X}$ gates were on the order of $10^{-3}$ and $10^{-4}$, respectively, and the readout error was on the order of $10^{-2}$.

In this study, we have chosen Trotter numbers so that the Trotter errors in the quantum-circuit simulation are acceptable for evaluating the magnetization and twist overlap.
Increasing the Trotter number improves the approximation accuracy, but also increases noise-induced errors in real-device computation.
Two-qubit gates contribute larger errors than single-qubit gates, and their number increases almost linearly with the Trotter number, regardless of the order of the Suzuki--Trotter decomposition used.
Thus, noise-induced errors are more sensitive to the Trotter number than the order of the Suzuki--Trotter decomposition.

\subsection{Difference in Effect of Errors}

As shown in Fig.~\ref{fig:w_mzz2r}, the twist overlap suffers from noise-induced errors more severely than the magnetization.
In the localized regime, the real-device computation results for $M_z$ are approximately $3/4$ of the exact time evolution result, whereas those of $|z|^2$ are less than $1/3$ of the exact time evolution result.
The decay ratio of the average and standard deviation is almost the same, suggesting that noise-induced errors are more prominent in the time average part than in the oscillating parts.
Considering that $\overline{M_z(t)}$ is estimated as Eq.~\eqref{eq:av_mz}, we can suppose that noise-induce errors reduce the dominant elements of $|c_k|^2|b_k|^2$ to $3/4$ of the expected value and that these dominant elements correspond to states with large $m_k$ values.
On the other hand, the $\overline{z(t)}$ of real-device computation results, which is estimated as Eq.~\eqref{eq:av_z2}, is much less than $(3/4)^2$ of the expected value.
The unexpectedly small values of $\overline{z(t)}$ imply that errors in the states with small $m_k$ values affect the twist overlap.

The error in $|z|^2$ is approximately three times larger than that in $M_z$.
As both the twist overlap and the magnetization are derived from measurements of individual qubits, this relationship remains independent of the noise strength.
If we manage to reduce the error in $M_z$ to around 3\%, the error in $|z|^2$ becomes approximately 10\%.
The reduction in error makes the twist overlap a potentially more informative quantity. 

\subsection{Information on Eigenstates}

Figure~\ref{fig:w_mzz2s}(c) illustrates that the disorder dependence of $|z|^2$ evaluated with the wavefunction after time evolution is similar to that evaluated with the eigenstates of the Hamiltonian.
The twist overlap $z_e$ evaluated with the eigenstates is described as
\begin{align}
 |z_e|^2 = \frac{1}{|B|}\sum_{k\in B}
\left|\langle\phi_k|\hat{U}_{\rm twist}|\phi_k\rangle\right|^2
\simeq \frac{1}{|B|}\sum_{k\in B} |b_k|^4,
\label{eq:ze}
\end{align}
where $B$ denotes the index set of the eigenstates used for the evaluation and $|B|$ is its size.
Equation~\eqref{eq:ze} indicates that $|z_e|^2$ approximately represents the simple average of $|b_k|^4$ in set $B$ in the localized regime.
In contrast, $\overline{|z(t)|^2}$, which is estimated as Eq.~\eqref{eq:av_z2} in the localized regime, corresponds to an average weighted with the overlap between eigenstates and the initial state.
In the localized regime, where $|b_k|^2\approx 1$ and $|c_0|^2\approx 1$ for a typical sample, $|z_e|^2 \simeq \overline{|z(t)|^2}$ on average.

In contrast, as shown in Fig.~\ref{fig:w_mzz2s}(d), the variance of $|z|^2$ is more significant for the twist overlap evaluated after time evolution than for that evaluated with eigenstates.
This increase in variance is due to the oscillating part of $|z(t)|^2$.
Since the oscillation amplitude is significant in some samples, the twist overlap can have various values at the end of time evolution, leading to a significant variance.

In the quantum-circuit simulation and real-device computation, the evaluation of the twist overlap is based on the measurement of each qubit.
It is not a straightforward summation of measured values ($s_j$); rather, it is expressed as an infinite series in terms of $s_j$. 
An alternative method for evaluating the twist overlap is to employ the Hadamard test between $|\psi(t)\rangle$ and $\hat{U}_{\rm twist}|\psi(t)\rangle$, which includes the measurement of only a single ancilla qubit.
This approach is applicable in quantum-circuit simulation as gate errors are excluded.
However, the Hadamard test involves control-$U$ operations, potentially resulting in notable two-qubit gate errors in real-device computations.

\section{Conclusions}

In this study, we investigated the method to detect MBL based on quantum dynamics, focusing on magnetization and twist overlap.
We examined the disorder dependence of these observables and analyzed their time evolution in the localized regime both theoretically and numerically.
By analyzing the time evolution, we gained insights into how noise-induced errors affect the observables.

The magnetization is simple and robust against noise.
The disorder dependence of the magnetization clearly shows the crossover between thermal and localized regimes.
The magnetization is evaluated from measured values of spin configurations ($s_j$), which directly correspond to the eigenvalues ($\sum_j s_j$) of the magnetization operator.
Furthermore, the magnetization has a large expectation value for the initial state, leading to a significant contribution of the initial state and small contributions of other states.
This type of observable is robust against noise because states with small expectation values do not significantly contribute to the observable.

On the other hand, the twist overlap is vulnerable to noise-induced errors, although the disorder dependence exhibits a weak signature of the crossover in real-device computation results.
The evaluation of the twist overlap relies on the measured values ($s_j$) through the phases ($u=\sum_j(\pi j/L)s_j$) of the eigenvalues ($\exp(iu)$) of the twist operator.
However, the information of phases is lost in the absolute value of the twist overlap, and that of eigenstates of the operator emerges.
Furthermore, every eigenstate of the Hamiltonian can contribute to the twist overlap, implying that noise-induced errors are not systematically reduced, unlike the magnetization case.
Despite being vulnerable to noise, this type of observable can provide richer information about the eigenstates of the Hamiltonian than other observables such as magnetization under a noiseless condition.

In conclusion, quantum simulations using a real noisy device can detect the crossover between thermal and localized regimes, provided that appropriate observables are chosen.
Some observables, such as the magnetization, are robust against noise-induced errors, whereas others, such as the twist overlap, are vulnerable.
The results of this study provide helpful insights into the selection of observables for probing MBL using a real quantum computer.


\begin{acknowledgments}
We acknowledge the use of IBM Quantum services for this work. 
The views expressed are those of the author and do not reflect the official policy or position of IBM or the IBM Quantum team.
This work was partially supported by the JSPS KAKENHI Grant Number JP23H04499.
\end{acknowledgments}

\bibliography{mbl.bib}

\end{document}